\documentclass[epj]{webofc}
\usepackage[varg]{txfonts}   % Web of Conferences font
\newcounter{comment}

{\refstepcounter{comment}%
\begin{quote}
\color{red}\sffamily\small$\blacksquare$ \textbf{\underline{Comment} $\sharp$\thecomment:}}%
{\end{quote}}

{\begin{quote}
\sffamily\small$\blacktriangleright$ \textbf{\underline{Reply} $\sharp$\thecomment:}}%
{\end{quote}}
\usepackage{multirow}

\woctitle{6\textsuperscript{th} International conference on
Physics Opportunities at Electron-Ion colliders (POETIC6)}

\newcommand{\xB}{x_{\rm B}}
\newcommand{\Q}{Q}
\newcommand{\GeV}{{\rm GeV}}
\newcommand{\GK}{\texttt{GK}}

\newcommand{\KM}[1]{\texttt{KM#1}}

\newcommand{\Asin}[2]{A^{\sin(#1\phi)}_{\rm #2}}
\newcommand{\Acos}[2]{A^{\cos(#1\phi)}_{\rm #2}}

\newcommand{\BSS}[1]{{\Sigma}^{\cos(#1\phi)}}
\newcommand{\BSD}[1]{{\Delta}^{\sin(#1\phi)}}

\begin{document}
\title{Description and interpretation of DVCS measurements%
\footnote{Joint proceedings contribution, based on two talks given by the authors at the
  6\textsuperscript{th} International conference on
Physics Opportunities at Electron-Ion colliders (POETIC6)}
}

\author{Kre\v{s}imir Kumeri\v{c}ki\inst{1}\fnsep\thanks{\email{kkumer@phy.hr} } \and
Dieter M\"uller\inst{2,}\inst{3}\fnsep\thanks{\email{dieter.mueller@irb.hr}}
      }

\institute{
Physics Department, University of Zagreb, %Bijeni\v{c}ka c. 32
10000 Zagreb, Croatia
\and
Theoretical Physics Division, Rudjer Bo{\v s}kovi{\'c} Institute, 10002 Zagreb, Croatia
\and
Department of Physics, University of Cape Town, 7701 Rondebosch, South Africa
}

\abstract{%
Having in mind the well-known limitations of certain models of
generalized parton distributions (GPDs),
we show that the allegedly universal GPDs, describing both
deeply virtual Compton scattering (DVCS) and
deeply virtual meson production (DVMP) data,
fail to describe measurements of deep inelastic scattering.
We present a new global DVCS fit that describes reasonably the world
data set, which includes now also new measurements from Hall A and CLAS
collaborations.  We also explicitly illustrate that Compton form factors (CFFs)
cannot be unambiguously extracted from photon electroproduction measurements
off unpolarized proton alone and we argue that it is more reliable to
interpret such measurements in terms of certain CFF combinations, where still
some care is needed in order to estimate the propagated error.
}
\maketitle
\section{Introduction}
\label{intro}
The deeply virtual Compton scattering (DVCS) process off a nucleon target is considered to be the theoretically cleanest process allowing access to generalized parton distributions (GPDs) from experimental measurements. Motivated by this, deeply virtual Compton scattering has been studied
in the last decade quite extensively by HERA and JLab collaborations. Thereby, cross section measurements on an unpolarized proton have been performed in collider experiments
H1 and ZEUS as well as now in fixed target experiments by the CLAS and Hall A collaborations, while HERMES provided an almost (over-)complete set of asymmetry measurements. On the other hand, considering possible photon and proton helicities,
there are twelve DVCS amplitudes where, loosely spoken, the four photon helicity conserved amplitudes or Compton form factors (CFFs) ${\cal H}, {\cal E}, \widetilde{\cal H}, \widetilde{\cal E}$ appear in the $1/\Q^2$ expansion as leading contributions. In the leading order and leading twist approximation these CFFs are given in terms of quark GPDs ${ H}, {E}, \widetilde{ H}, \widetilde{E}$,
\begin{eqnarray}
\label{eq:CFF}
{\cal H} (x_B,t,{\cal Q}^2) &\!\!\!\stackrel{\rm LO}{=}\!\!\!&
  \int_{-1}^1\!dx\, \left[\frac{1}{\xi-x- i \epsilon}-\frac{1}{\xi+x-i \epsilon} \right]
 \sum_{q=u,d,s,\cdots} e_q^2 H^q (x,\xi,t,{\cal Q}^2)\,,
\end{eqnarray}
where $\xi \approx x_B/(2-x_B)$, $x_B$ is Bjorken's scaling variable, $t$ is the square of the momentum transfer, and the
factorization scale has been equated to the (negative) photon virtuality $q^2=-{\cal Q}^2$.
The proton helicity flip CFFs $\cal E$ and $\widetilde{\cal E}$ might be kinematically suppressed in the DVCS kinematics, \emph{e.g.}, by $t/4M^2$ ($M$ being the target mass); however, $\widetilde{\cal E}$ might also be enhanced at small $-t$ by the presence of the pion pole. The rather analogous set of four longitudinal photon helicity CFFs is formally suppressed by $1/\Q$.
The remaining four transverse helicity photon flip CFFs appear in the partonic description at next-to-leading order accuracy via the gluon transversity GPDs which might also be to a larger extent contaminated by $1/\Q^2$ power suppressed contributions. Apart from the desire to access GPDs in the deeply virtual region where the photon virtuality is considered to be large, we require
$\Q^2 \gtrsim 1.5 \GeV^2$, and the momentum transfer $-t$ with $-t/\Q^2 \lesssim 1/4$ is small, virtual Compton scattering has its own interest. A unifying framework was proposed in Ref.\ \cite{Belitsky:2012ch} in which also the low energy limit, \emph{i.e.}, the access to generalized polarizabilities was discussed.

The DVCS process is measurable in the
leptoproduction of a photon on a nucleon (or nucleus) target and it interferes with the Bethe-Heitler process, which is parameterized in terms of the electro-magnetic form factors. For a transversally polarized target the differential cross section is five-fold and is considered as function of $\xB$, $t$, $\Q^2$, the azimuthal angle $\phi$ between the lepton and reaction plane, and the azimuthal angle $\varphi$ of the transverse components of the polarization vector. Otherwise, the cross section is four-fold and for a massless electron reads
\begin{equation}
\label{dsigma}
\frac{d^4\sigma}{d\xB  d\Q^2 dt  d\phi} = \frac{\alpha^3 \xB y^2}{8 \pi \Q^4 \sqrt{1+\epsilon^2}} \left|\frac{{\cal T}}{e^3}\right|^2,
\end{equation}
where $\alpha=e^2/4\pi \approx 1/137$ is the electromagnetic fine structure constant, $y=\Q^2/\xB\left(s-M^2\right)$ with $s=(p_1+q_1)^2$, and
$\epsilon^2 = 4\xB^2 M^2/\Q^2$ is an abbreviation.
 The electroproduction amplitude, entering as  square of its absolute value into the cross section,
\begin{equation}
|{\cal T}|^2 = |{\cal T}^{\rm BH}|^2 + {\cal I} + |{\cal T}^{\rm VCS}|^2,
\quad
{\cal I} = 2{\Re}{\rm e}{\cal T}^{\rm BH} {\cal T}^{\dagger\, {\rm VCS}}
\end{equation}
is the sum of the charge-odd Bethe-Heitler ${\cal T}^{\rm BH}$ amplitude and the charge-even virtual Compton scattering amplitude ${\cal T}^{\rm VCS}$ amplitude.
The three different contributions might be decomposed in Fourier harmonics, \emph{e.g.}, for an unpolarized nucleon
\begin{subequations}
\label{FH-decomposition}
\begin{equation}
\label{BHsquared}
 |{\cal T}^{\rm BH}|^2
=
\frac{e^6}{\xB^2 y^2 (1+\epsilon^2)^2 t\, {\cal P}_1 (\phi) {\cal P}_2 (\phi)} \sum_{n=0}^2 c^{\rm BH}_{n, {\rm unp}}\cos(n\phi)\,,
\end{equation}
where the product of (rescaled)  electron propagators  $1/{\cal P}_1 (\phi) {\cal P}_2 (\phi)$ implies an additional $\phi$ dependence.
The coefficients $c^{\rm BH}_{n, {\rm unp}}$ depend on the electromagnetic proton form factors for which we use the Kelly parametrization \cite{Kelly:2004hm}.
For the interference term we have
\begin{eqnarray}
\label{InterferenceTerm}
{\cal I}
&\!\!\!=\!\!\!&
\frac{\pm e^6}{\xB y^3 t {\cal P}_1 (\phi) {\cal P}_2 (\phi)}
\Bigg[
\sum_{n = 0}^3
c_{n, {\rm unp}}^{\rm I}\, \cos(n \phi)
+
\sum_{n = 1}^2  s_{n, {\rm unp}}^{\rm I}\, \sin(n \phi)
\Bigg] ,
\end{eqnarray}
where the $+$ ($-$) sign refers to electron (positron) scattering and
\begin{equation}
\label{VCSquared}
 |{\cal T}^{\rm VCS}|^2
=
\frac{e^6}{y^2 {\Q}^2}\left[  \sum_{n=0}^2 c^{\rm VCS}_{n, {\rm unp}}\cos(n\phi) + s^{\rm VCS}_{1,{\rm unp}}\sin(\phi) \right],
\end{equation}
\end{subequations}
for the squared DVCS amplitude. The harmonics of the interference term are given in terms of linear CFF combinations
\begin{eqnarray}
\label{C_unp}
{\cal C}^{\cal I}_{\rm unp} &\!\!\!=\!\!\!& F_1 {\cal H}+ \xi (F_1+F_2) \widetilde {\cal H} -\frac{t}{4 M^2} F_2 {\cal E}\,,
\\
\label{C_LP}
{\cal C}^{\cal I}_{\rm LP} &\!\!\!\approx\!\!\!&  F_1 \widetilde{\cal H} + \xi (F_1+F_2) {\cal H}
-
\left(\xi F_1 + \frac{t}{4M^2}F_2
\right)\xi \widetilde{\cal E}\,.
\end{eqnarray}
while those of the squared DVCS term are given in terms of bilinear (linear
in both CFFs and their complex conjugates) combinations.
%\begin{eqnarray}
%\end{eqnarray}
Projecting first the experimental data onto (weighted) harmonics, see \cite{Belitsky:2001ns}, is a standard filter procedure we utilize first in both global GPD model fitting and local CFF extraction.

The main emphasis of the present contribution is to spell out some phenomenological inconsistencies in the description of DVCS data which are not very much
clarified in the literature.
In Sec.\ \ref{sec:model_vs_fit} we show that universality of the GPD framework is violated in a certain model class and we provide a description of the world data DVCS set that is obtained by global fitting. In Sec.\ \ref{sec:local} we spell out that the interpretation of some results that arise from so-called quasi model-independent extraction of CFFs from CLAS cross section measurements might be misleading. Finally, we conclude.

\section{Model predictions versus global fitting}
\label{sec:model_vs_fit}

Based on Regge theory arguments, one might assume that in collider experiments only one CFF contributes, giving access to the unpolarized quark GPD  $H^q$ combination, \emph{e.g.}, in LO approximation by means of the convolution formula (\ref{eq:CFF}).
The situation is different for fixed target kinematics, in particular for the JLab experiments with a $6\, {\rm GeV}$ electron beam and unpolarized or longitudinally polarized proton target. Additionally to model predictions, global fitting attempts were undertaken to extract  the values of CFFs. It is clear from the outset that also in the case of a complete or overcomplete measurement the extraction of such information is a challenging task that can only be resolved within some physical or model-motivated assumptions. The common assumption most authors adopt is the dominance of leading twist (LT) in a perturbative leading order (LO) description, \emph{i.e.}, the number of complex-valued Compton form factors
$$
{\cal F} \in \{{\cal H},{\cal E},\widetilde{\cal H},\widetilde{\cal E},\cdots\}
$$
reduces from twelve to four. This restriction might be not justified outside the DVCS kinematics, \emph{e.g.}, $-t \sim \Q^2/2$ or so.

The new CLAS beam spin asymmetry $A_{\rm LU}$ and longitudinally polarized target spin asymmetries $A_{\rm UL}$  and $A_{\rm LL}$ are measured in five $\{\xB,\Q^2\}$--bins versus $t$ \cite{Pisano:2015iqa}, where four bins cover DVCS kinematics
\begin{eqnarray}
\{0.185,1.63\, {\rm GeV}^2 \},\;  \{0.245, 1.79\, {\rm GeV}^2 \},\;
\{0.245, 2.12\,{\rm GeV}^2  \},\;   \{0.335, 2.78\, {\rm GeV}^2 \}.
\label{CLAS15_bin-1}
\end{eqnarray}

The Hall A collaboration provided high precision cross section measurements with a polarized electron beam and unpolarized proton target \cite{Munoz_Camacho:2006hx}, which were superseded more recently for five $-t/\GeV^2\in \{0.17,0.23,0.28,0.33,0.37\}$ values within five kinematical settings \cite{Defurne:2015kxq},
\begin{eqnarray}
{\rm Kin1:} && 0.345 \le  \xB  \le 0.399\,,\;\; 1.453\le \Q^2/\GeV^2 \le 1.633\,,\quad
\nonumber\\
{\rm Kin2:} && 0.343 \le  \xB  \le 0.381\,,\;\; 1.820\le \Q^2/\GeV^2 \le 1.999\,,\quad
\nonumber\\
{\rm Kin3:} && 0.345 \le  \xB  \le 0.373\,,\;\; 2.218\le \Q^2/\GeV^2 \le 2.375\,,\quad
\nonumber\\
{\rm KinX2:} && 0.378 \le \xB \le  0.401\,, \;\;2.012\le \Q^2/\GeV^2 \le 2.091\,,\quad
\nonumber\\
{\rm KinX3:} && 0.336 \le \xB \le  0.342\,,\;\; 2.161\le \Q^2/\GeV^2 \le 2.193\,,\quad
%\nonumber\\
\label{HallA-KINs}
\end{eqnarray}
where last two settings have subset of the same data re-binned for measuring
the $x_B$-dependence.
Somewhat less precise measurements of cross-section, but covering
much larger kinematic space
\begin{equation}
0.1 \le \xB \le  0.58\,,\;\; 1\le \Q^2/\GeV^2 \le 4.6\,,\quad
0.17\le -t/\GeV^2 \le 0.37 \,,
\label{CLAS15-KIN}
\end{equation}
were published by CLAS \cite{Jo:2015ema}.

\subsection{Model predictions}
\label{sec:model}

It is known since more than a decade that certain GPD models such as Radyushkin`s double distribution ansatz (RDDA) or the leading SO(3)-partial wave expansion of conformal GPD moments fail to globally describe the DVCS data at leading twist and leading order of perturbation theory.
While the failure of the latter model is well-accepted, problems of the former one
are often forgotten or hidden in the literature, see, e. g., Ref.~\cite{Kroll:2012sm},
where a good description of DVCS data by RDDA-based model has been reported.

It is rather popular to decorate the RDDA with $t$-dependence in such a manner,
\begin{eqnarray}
H^q(x,\eta,t) &\!\!\!\!=\!\!\!\!& \int_{0}^1\!dy \int_{-1+y}^{1-y}\!dz\, \delta(x-y-z \eta)
%\\
%&&%\phantom{\int_{0}^1\!dy \int_{-1+y}^{1-y}}\times
\frac{q(y,t)}{1-y}\, \Pi\!\left(\!\frac{z}{1-y}\!\right) +
\theta(|x| \le |\eta|) D\!\left(\!\frac{x}{|\eta|},t\!\right),
%\nonumber
\label{DD-H}
\end{eqnarray}
that the $t$-dependence is entirely contained in the non-forward parton density $q(x,t)$. The model bias in such a parametrization is
that the profile function is $t$-independent and, thus, for valence quarks it can be adjusted from form factor measurements where one relies on
the functional ansatz. To complete polynomiality in the charge even sector, one might add a so-called $D$-term. Such an ansatz is the basis of older
models, where partially a factorized $t$-dependence was utilized, the \GK{} model \cite{Goloskokov:2007nt}, and it is also implemented in the VGG code \cite{Vanderhaeghen:1999xj}.

\begin{figure}
\begin{center}
\includegraphics[scale=0.52]{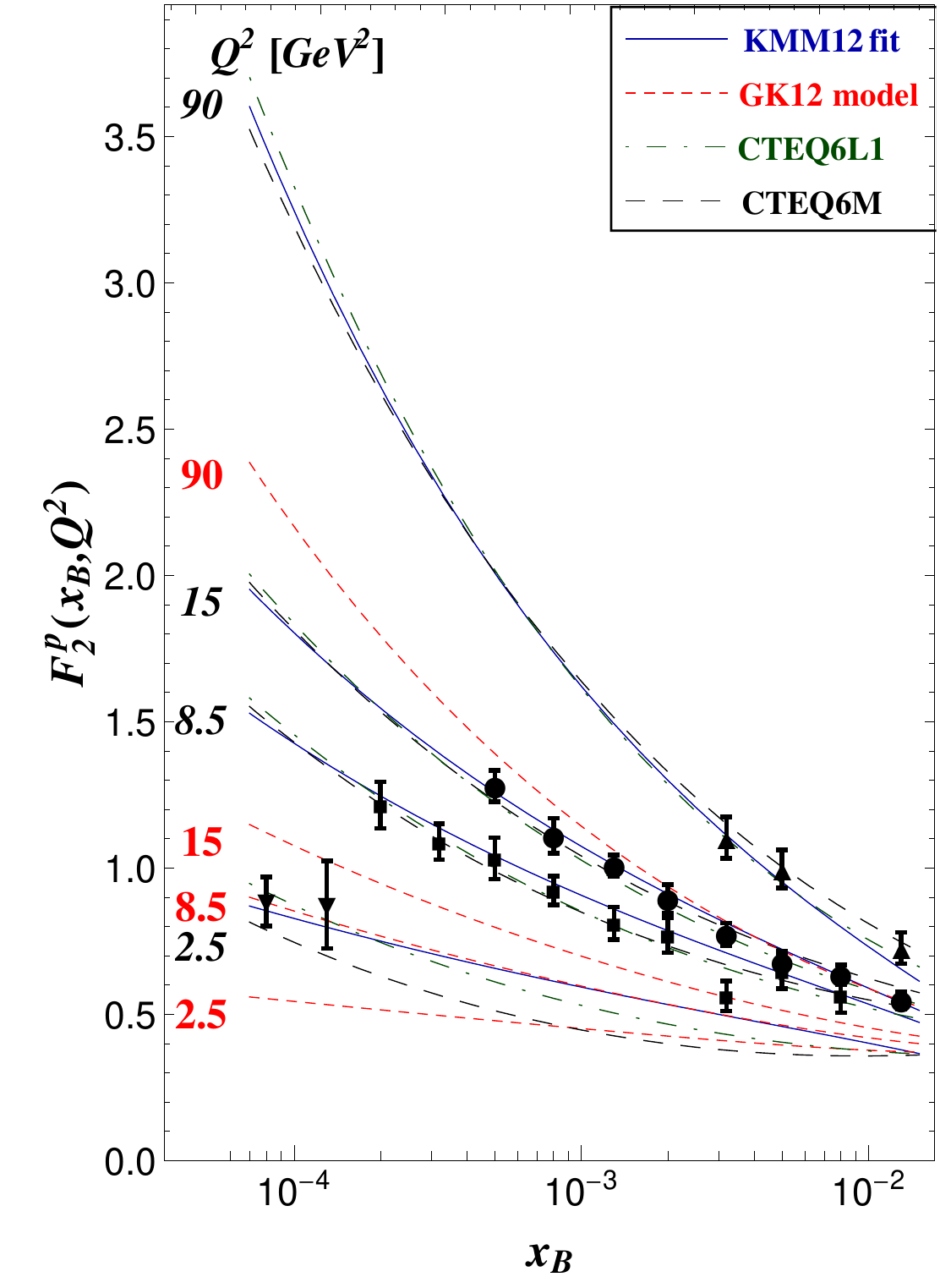}
\includegraphics[scale=0.55]{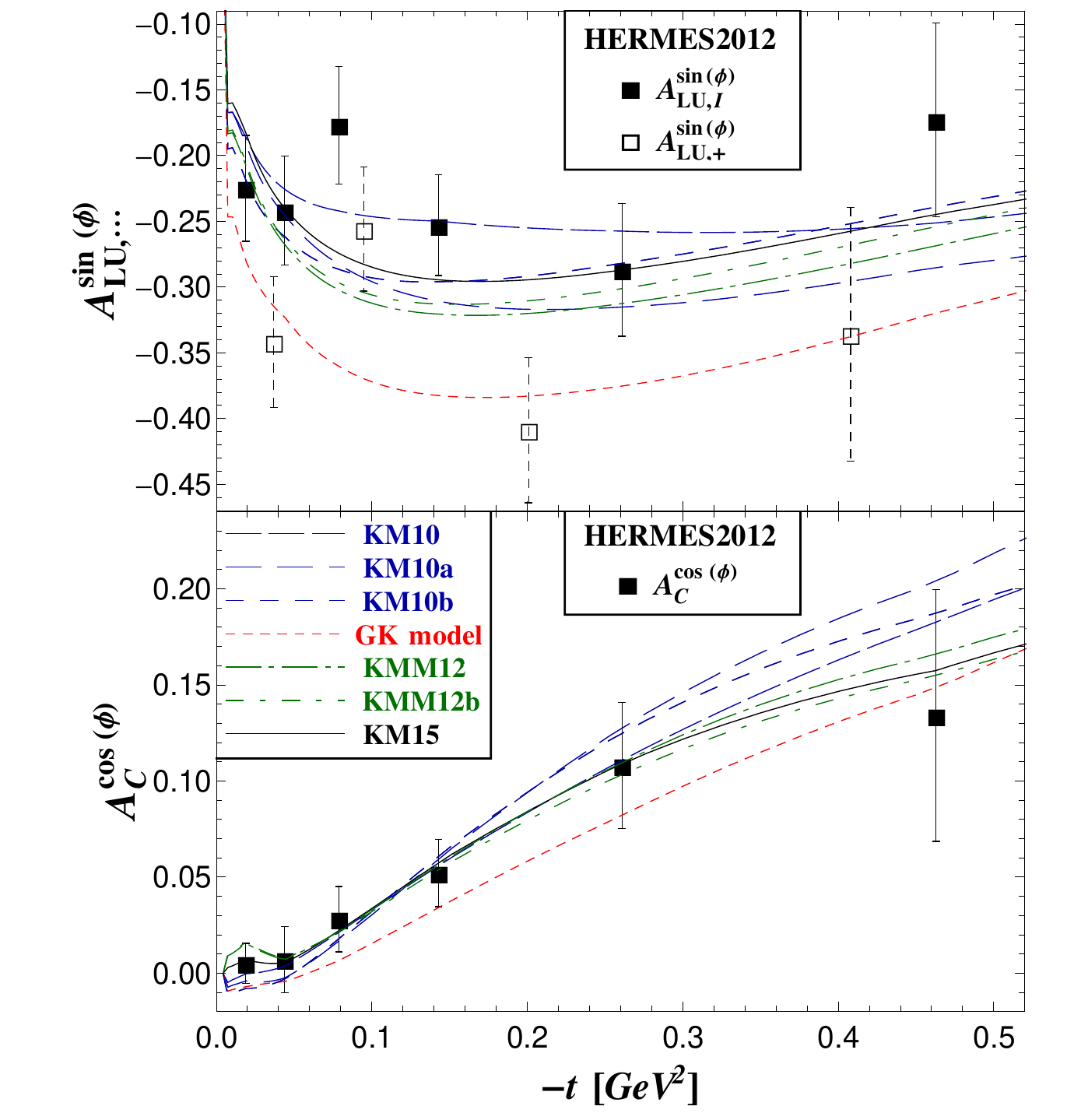}
\end{center}
\caption{Left: The  DIS structure function $F_2^p$  measurements \cite{Aid:1996au}  are confronted with the \GK\ model (short--dashed), based on the next-to-leading order PDF parametrization CTEQ6M (dashed),  the leading order PDF parametrization CTEQ6L1 (dot--dashed)  as well as our DIS fit (solid) from Ref.\ \cite{Kumericki:2009uq}. Right:
Beam spin (upper panel) and beam charge (lower panel) asymmetry from the HERMES collaboration:  measurements of $\Asin{}{LU,+}$ with the recoil detector (empty squares) \cite{Airapetian:2012pg} and
obtained from  measurements of $\Asin{}{LU,I}$ and $\Acos{}{C}$ within the missing mass technique (filled squares)  \cite{Airapetian:2009aa} versus  the \GK\ model prediction (short--dashed) and various fits, \KM{10}\ (long-dashed),  \KM{10a}\ (dashed), \KM{10b}\ (shorter dashed),  \KM{M12}\ (dash--dotted),  \KM{M12b}\ (shorter dash--dotted), and \KM{15} (solid), to the HERMES data (filled squares)  as solid curves.
\label{fig:F2}
}
\end{figure}

The phenomenological problem with
such models is that the skewness effect at small $\xB$ is too large and so their predictions substantially overshoot the DVCS cross section measurements in collider kinematics, as pointed out for some time already in Ref.\ \cite{Freund:2002qf}. Thus, it is at first surprising that the \GK{} model correctly describes the H1 and ZEUS DVCS data. However, the reason for that is the reduction to $n_F=3$ of the flavour content of the PDF parametrization, adopted from CTEQ6 \cite{Pumplin:2002vw}, see convolution formula (\ref{eq:CFF}).
Such exclusion of the charm quark contribution from the model leads then to the large
underestimate of the deep inelastic structure function $F_2(\xB,Q^2)$.
This is displayed by the short dashed line in the left panel of Fig.~\ref{fig:F2}
to be compared with the LO  CTEQ6L1 (dash-dotted), NLO CTEQ6M
(dashed), and our LO parametrization (solid).  Hence, we conclude that the
universality of the GPD framework in the \GK{} model (and other GPD models based on RDDA) is not implemented.

On the other hand the \GK{} model describes the asymmetry measurements of HERMES that were obtained within the missing mass technique, except for the beam spin asymmetry $\Asin{1}{LU}$, which is a key observable, compare full squares with short dashed line in the upper right panel of Fig.\ \ref{fig:F2}. On the other hand the recoil detector data of the HERMES collaboration (empty squares), which are consistent with the missing mass technique data, are described rather well by the \GK{} model. Unfortunately, the two different data sets are not combined and the missing mass technique data suffer from a larger experimental uncertainty. As we will see in the next section the \GK{} model quantitatively overestimates the beam spin asymmetry and the beam spin dependent cross section of JLab measurements.  \subsection{Predictions of new JLab results and a new global fit}
To fit globally the world data set of DVCS measurements from HERA and JLab
collaborations, we employ a rather simple hybrid model.
It is based on a double partial wave expansion for the sea-quark and gluon GPD,
including the LO QCD evolution, and a dispersion integral representation
for the valence quark content that does not evolve.
Details of the model can be found in \cite{Kumericki:2009uq}, and last
published incarnation of the model (\KM{M12}) \cite{Kumericki:2013br}
was obtained by a standard
least-squares fitting procedure to the statistically independent set of 95 DVCS
data points\footnote{Actually, only 92 points were independent; we took
  all 6 CLAS $A_{UL}^{\sin\phi}$ measurements \cite{Chen:2006na},
  where only 3 are independent.
  This is corrected in the new \KM{15} fit.}.
 Total $\chi^2/n_{\rm d.o.f.}$ was 123.5/80, where most problematic
data was (i) HERMES data with longitudinally polarized target (and to some
extent beam), and (ii) the very precise Hall A cross-section data (especially
$t$-dependence of the first cosine harmonic).

On Table~\ref{tab:pulls} we list all data used in fits, as well as
$\chi^2/n_{\rm pts}$ values for particular datasets, for the \KM{M12}
model, and for the new \KM{15} model to be discussed below. To give additional
information about compatibility of the fit with the data, we list
also the values of the ``pull'' of each dataset, defined for
dataset consisting of $N$ measurements as
\begin{equation}
    \text{pull} \equiv \frac{1}{\sqrt{N}} \sum_{i=1}^N
    \frac{ f(x_i)- y_i }{\Delta y_i} \;,
    \label{eq:pull}
\end{equation}
where $f(x_i)$ is model value for a given kinematics $x_i$, while
$y_i$ and $\Delta y_i$ are the corresponding experimental value and
uncertainty, respectively. For a perfect model,
pull (\ref{eq:pull}) would be a random variable distributed according to
a Gaussian distribution with zero mean and unit standard deviation.

\begin{table}
\caption{
    Observables used for \KM{M12} and \KM{15} fits, together with
  corresponding values of $\chi^2/n_{\rm pts}$ and ``pulls''
  (\protect\ref{eq:pull}).
}
\label{tab:pulls}
\centering
\renewcommand{\arraystretch}{1.2}
\begin{tabular}{ccccccccc}
\hline\noalign{\smallskip}
\multirow{2}{*}{Collaboration} & \multirow{2}{*}{Observable}& \multirow{2}{*}{Ref.} &
\multirow{2}{*}{$n_{\rm pts}$}  & \multicolumn{2}{c}{\KM{M12}} & &
\multicolumn{2}{c}{\KM{15}} \\
\cline{5-6}\cline{8-9}\noalign{\smallskip}
& & & & $\chi^2/n_{\rm pts}$ & pull & & $\chi^2/n_{\rm pts}$ & pull \\
\noalign{\smallskip}\hline
%%% BEGIN ROWS
ZEUS & $\sigma_{\rm DVCS}$ & \cite{Chekanov:2008vy,Chekanov:2003ya} & 11 &  0.49 & -1.76 & &  0.51 & -1.74 \\
ZEUS,H1 & $d\sigma_{\rm DVCS}/dt$ & \cite{Aaron:2009ac,Chekanov:2008vy,Aktas:2005ty} & 24 &  0.97 &  0.85 & &  1.04 &  1.37 \\
HERMES & $A_{\rm C}^{\cos 0\phi}$ & \cite{Airapetian:2012mq} & 6 &  1.31 &  0.49 & &  1.24 &  0.29 \\
HERMES & $A_{\rm C}^{\cos \phi}$ & \cite{Airapetian:2012mq} & 6 &  0.24 & -0.56 & &  0.07 & -0.20 \\
HERMES & $A_{\rm LU,I}^{\sin \phi}$ & \cite{Airapetian:2012mq} & 6 &  2.08 & -2.52 & &  1.34 & -1.28 \\
CLAS & $A_{\rm LU}^{\sin \phi}$ & \cite{Girod:2007aa} & 4 &  1.28 &  2.09 & &       &       \\
CLAS & $A_{\rm LU}^{\sin \phi}$ & \cite{Gavalian:2008aa,Pisano:2015iqa} & 13 &       &       & &  1.24 &  0.63 \\
CLAS & $\Delta\sigma^{\sin\phi,w}$ & \cite{Jo:2015ema} & 48 &       &       & &  0.41 & -1.66 \\
CLAS & $d\sigma^{\cos 0\phi,w}$ & \cite{Jo:2015ema} & 48 &       &       & &  0.16 & -0.21 \\
CLAS & $d\sigma^{\cos\phi,w}$ & \cite{Jo:2015ema} & 48 &       &       & &  1.16 &  6.36 \\
Hall A & $\Delta\sigma^{\sin\phi,w}$ & \cite{Munoz_Camacho:2006hx} & 12 &  1.06 & -2.55 & &       &       \\
Hall A & $d\sigma^{\cos 0\phi,w}$ & \cite{Munoz_Camacho:2006hx} & 4 &  1.21 &  2.14 & &       &       \\
Hall A & $d\sigma^{\cos\phi,w}$ & \cite{Munoz_Camacho:2006hx} & 4 &  3.49 & -0.26 & &       &       \\
Hall A & $\Delta\sigma^{\sin\phi,w}$ & \cite{Defurne:2015kxq} & 15 &       &       & &  0.81 & -2.84 \\
Hall A & $d\sigma^{\cos 0\phi,w}$ & \cite{Defurne:2015kxq} & 10 &       &       & &  0.40 &  0.92 \\
Hall A & $d\sigma^{\cos\phi,w}$ & \cite{Defurne:2015kxq} & 10 &       &       & &  2.52 & -2.42 \\
HERMES,CLAS & $A_{\rm UL}^{\sin \phi}$ & \cite{Airapetian:2010ab,Chen:2006na} & 10 &  1.90 & -1.89 & &  1.10 & -1.94 \\
HERMES & $A_{\rm LL}^{\cos 0 \phi}$ & \cite{Airapetian:2010ab} & 4 &  3.44 &  2.17 & &  3.19 &  1.99 \\
HERMES & $A_{\rm UT,I}^{\sin(\phi-\phi_S) \cos \phi}$ & \cite{Airapetian:2008aa} & 4 &  0.90 &  0.61 & &  0.90 &  0.71 \\
CLAS & $A_{\rm UL}^{\sin \phi}$ & \cite{Pisano:2015iqa} & 10 &       &       & &  0.76 &  0.38 \\
CLAS & $A_{\rm LL}^{\cos 0 \phi}$ & \cite{Pisano:2015iqa} & 10 &       &       & &  0.50 & -0.22 \\
CLAS & $A_{\rm LL}^{\cos  \phi}$ & \cite{Pisano:2015iqa} & 10 &       &       & &  1.54 &  2.40 \\
%%% END ROWS
\noalign{\smallskip}\hline
%\textbf{}
\end{tabular}
\renewcommand{\arraystretch}{1.}
\end{table}

\begin{table}
\caption{\label{tab:ps-KM15}\small Valence (top) and sea quark (bottom) related
hybrid model parameters, as obtained by the global DVCS fit \KM{15} to
290 data points with $\chi^2/{\rm d.o.f.} = 240./275$. For
detailed description of the model and parameters see
Refs.~\cite{Kumericki:2013br,Kumericki:2009uq}. Parameters of the
old \KM{M12} fit \cite{Kumericki:2013br} are also given for comparison.}
\centering
\renewcommand{\arraystretch}{1.2}
\begin{tabular}{ccccccccccc}
&$M^{\rm val}$ &  $r^{\rm val}$ &  $b^{\rm val}$ &  $C$ &  $M_C$ &
$\tilde{M}^{\rm val}$ &  $\tilde{r}^{\rm val}$ &  $\tilde{b}^{\rm val}$ &  $r_{\pi}$ &  $M_{\pi}$ \\
\KM{15} &0.789 & 0.918 & 0.4 & 2.768 & 1.204 & 3.993 & 0.881 & 0.4 & 2.646 & 4.\\[1pt]

\KM{M12} &0.95 & 1.12 & 0.4 & 1. & 2.08 & 3.52 & 1.3 & 0.4 & 3.84 & 4. \\[1pt] \hline \\[-14pt]
&$(M^{\rm sea})^2$ &  $s_2^{\rm sea}$ &  $s_4^{\rm sea}$ & $s_2^{\rm G}$  & $s_4^{\rm G}$ & & & & & \\
\KM{15} & 0.482 & 1.071 & -0.366 & -2.991 & 0.905 & & & & &  \\
\KM{M12} &0.46 & 0.31 & -0.14 & -2.77 & 0.94  & & & & & \\ \hline
\end{tabular}
\renewcommand{\arraystretch}{1.}
\end{table}

\begin{figure}
\includegraphics[scale=0.4]{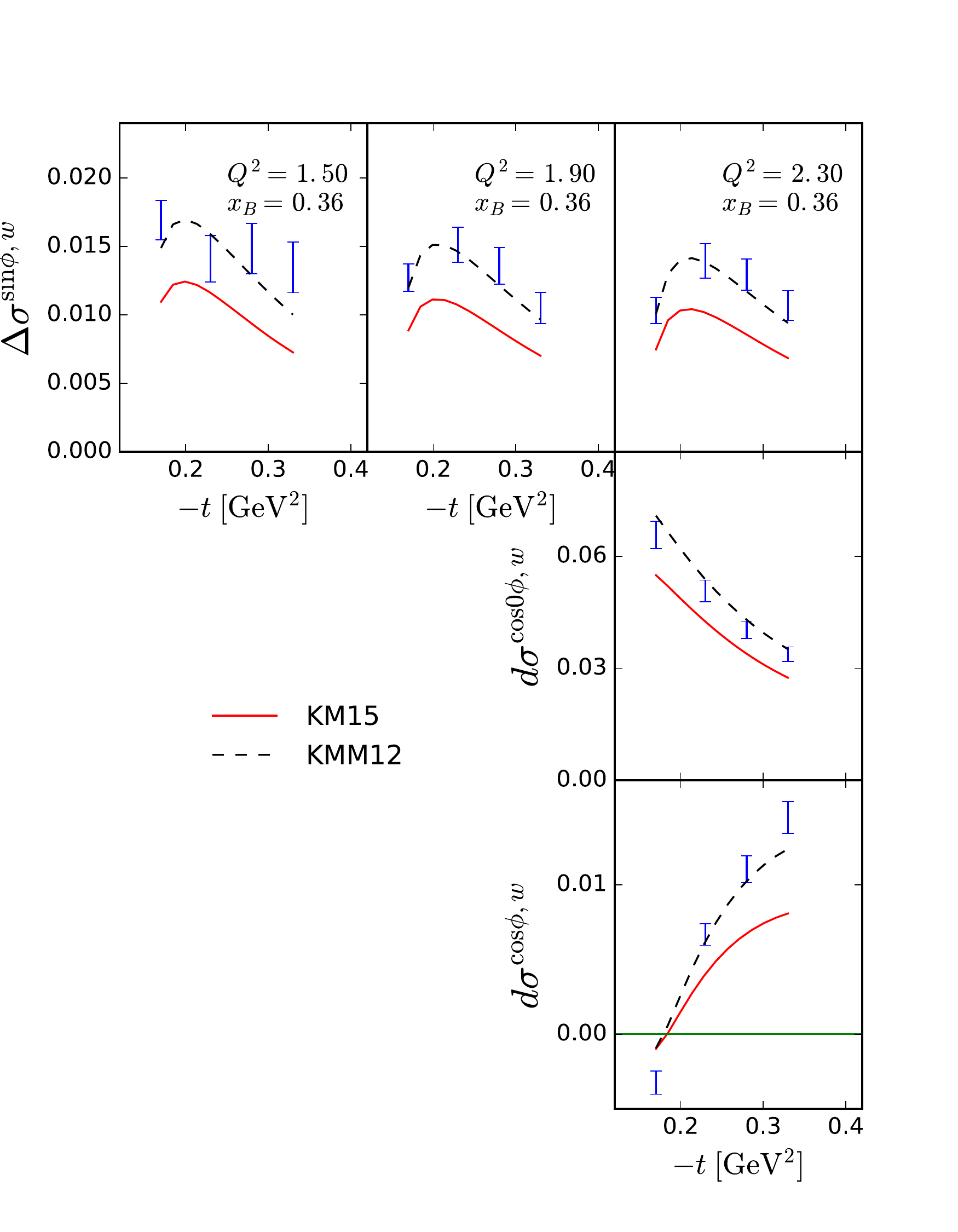}%
\includegraphics[scale=0.4]{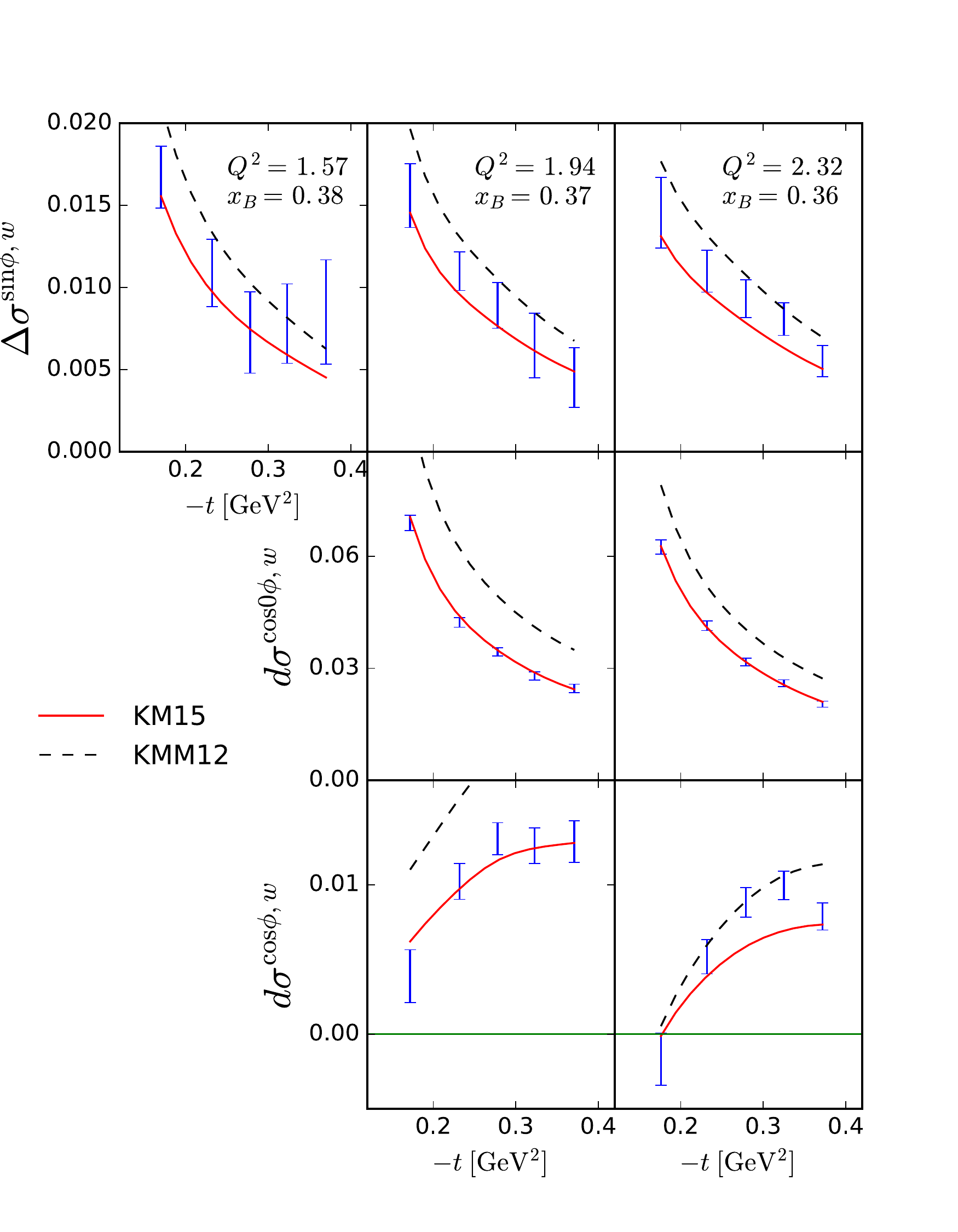}
\caption{\label{fig:HallA} (Weighted) Fourier harmonics
of Hall A 2006 \cite{Munoz_Camacho:2006hx} (left) and 2015
\cite{Defurne:2015kxq} (right) cross-section data, together
with \KM{M12} (dashed) and \KM{15} (solid) models.
}
\end{figure}

Adding 2015 JLab data to the world set, and refitting the same model, results
in a model parameter values listed on Table~\ref{tab:ps-KM15}.
The key difference is brought by using 2015 Hall A cross section data instead
of superseded 2006 one. Old and new Hall A data, and \KM{M12} and \KM{15}
models are confronted on Fig.~\ref{fig:HallA}. One notices that the cross-section
went down a bit and the strong
$t$-dependence of the first cosine harmonic ($d\sigma^{\cos\phi,w}$, lowest
panels) seen in 2006 data is significantly milder now.
All this made possible the decrease of skewness parameters
$r^{\rm val}$ and $\tilde{r}^{\rm val}$, {\it i.e.}, the decrease of
$\mathcal{H}$ and $\tilde{\mathcal{H}}$, as well as the decrease of
the pion pole contribution, thus improving the fit of most of the
other observables, see Table.~\ref{tab:pulls}.
In particular, description of HERMES $A_{\rm LU,I}^{\sin\phi}$ is significantly
improved, and overshooting (large positive pull) of
CLAS $A_{\rm LU}^{\sin\phi}$ is not present any more.

Looking at longitudinally polarized target data, $\chi^2$ of
$A_{\rm UL}^{\sin\phi}$ is also improved, while
very large $\chi^2$ contribution of HERMES
$A_{\rm LL}^{\cos 0\phi}$ is mostly due to a single outlier point so
we don't consider this a problem.
First cosine harmonic of the cross section remains the most problematic
observable to describe. As discussed above, in the Hall A case the tension
is relaxed but still exists, and although $\chi^2$ value for new
CLAS measurement is fine, the pull of 6.36 is unacceptably large,
showing consistent overshooting of the data.

\section{Local  versus  global extraction of CFFs}
\label{sec:local}

It is a standard problem in the phenomenology of scattering processes that if only an incomplete measurement is performed one can extract only certain combinations of squared helicity amplitudes at given kinematics.  In the photon leptoproduction the interference between Bethe-Heitler bremsstrahlung process and deeply virtual Compton scattering offers  various possibilities to access CFFs. To utilize the azimuthal angular dependence in the separation procedure, it is of advantage to perform first a Fourier analysis of the azimuthal angular dependence. In the following it is assumed that the eight sub-CFFs,
\begin{equation}
{\cal F} \in
\{\Im{\rm m}{\cal H},\Im{\rm m}{\cal E},\Im{\rm m}\widetilde{\cal H},\Im{\rm m}\widetilde{\cal E},\Re{\rm e}{\cal H},\Re{\rm e}{\cal E},\Re{\rm e}\widetilde{\cal H},\Re{\rm e}\widetilde{\cal E} \}\,,
\label{cff-set_LOLT}
\end{equation}
appearing in LO and LT approximation (\ref{eq:CFF}) or, equivalently, in the hand-bag approach, dominate the twist-two associated first and zeroth  asymmetry harmonics, where the latter one might be more contaminated by higher twist contributions than the former ones. Various methods for local extraction have been tested --- we will discuss here only two: map of random numbers and a so-called quasi model-independent fitter code that is based on reference GPD models.

\subsection{Mapping of random numbers}
\label{sec:mapping}

The  mathematical problem to extract CFFs locally, \emph{i.e.}, at fixed values of $\xB$, $\Q^2$, and $t$, from experimental measurements, can be simply understood as a map from the space of observables into the space of CFFs \cite{Belitsky:2001ns}. For asymmetry measurements with both kinds of leptons available this map of random numbers has been described in Ref.\ \cite{Kumericki:2013br}. For cross section measurements with a polarized electron beam and/or polarized nucleon target it can be abstractly formulated as a set of linear and quadratic equations,
\begin{eqnarray}
\label{F2A-gen}
\mbox{\boldmath $X$} &\!\!\!=\!\!\!& \mbox{\boldmath $b$}(F_1,F_2|\mbox{\boldmath ${\cal G}$}) +  \mbox{\boldmath $c$}^{-1}(F_1,F_2;\mbox{\boldmath ${\cal G}$}) \cdot \mbox{\boldmath $\cal F$}  + \mbox{\boldmath $\cal F$}^\dag\cdot \mbox{\boldmath $d$} \cdot \mbox{\boldmath $\cal F$} \,,
\quad
\mbox{\boldmath $X$} = \left(\begin{array}{c}
                               X_{\rm unp} \\
                               \vdots\\
                               \vdots\\
                                X_{h_n S_m}
                             \end{array}
\right),
\quad
\mbox{\boldmath $\cal F$} = \left(\begin{array}{c}
                               \Im{\rm m}H \\
                               \vdots\\
                               \Re{\rm e}H \\
                               \vdots
                             \end{array}
\right),
\end{eqnarray}
where the $n$-dimensional vector $\mbox{\boldmath $X$}$ contains the experimental measurements of (weighted) harmonics at given kinematics, $\mbox{\boldmath $\cal F$}$ is a $n$-dimensional vector of sub-CFFs (\ref{cff-set_LOLT}) under the assumption that the remaining $24-n$ dimensional sub-CFF vector $\mbox{\boldmath ${\cal G}$}$ can be considered as fixed. The constant vector
$$\mbox{\boldmath $b$}(F_1,F_2;\mbox{\boldmath ${\cal G}$})= \mbox{\boldmath $b$}^{\rm BH}(F_1,F_2)+
\mbox{\boldmath $b$}^{\rm INT}(F_1,F_2)\cdot\mbox{\boldmath ${\cal G}$}+\mbox{\boldmath ${\cal G}$}^\dag\cdot\mbox{\boldmath $b$}^{\rm VCS}\cdot\mbox{\boldmath ${\cal G}$}$$
contains contributions from the Bethe-Heitler term, interference term, and/or squared VCS amplitude, while the linear coefficient matrix $\mbox{\boldmath $c$}^{-1}(F_1,F_2;\mbox{\boldmath ${\cal G}$})$ arises from the interference term and/or squared VCS amplitude.  The (multi--valued) solution of the (quadratic) equations can be found by diagonalization.

Another local extraction method relies on the finding that apart from some power suppressed contributions, which are considered small, only certain linear and bilinear CFF combinations enter the interference term and the squared VCS amplitude, respectively. Strictly spoken, for each kinematical point we have different functional forms of these combinations which also depend on the center-of-mass energy or energy loss variable $y$. However, in the leading twist approximation the center-of-mass energy dependence approximately
separates and Eq.\ (\ref{F2A-gen}) might be written as
\begin{eqnarray}
\label{X2C-gen}
\mbox{\boldmath $X$} &\!\!\!\approx\!\!\!& \mbox{\boldmath $b$}(F_1,F_2) +  \mbox{\boldmath $k$}^{-1}\cdot
\mbox{\boldmath ${\cal C}$}({\cal F}|F_1,F_2)  +  \mbox{\boldmath $K$}\cdot
\mbox{\boldmath ${\cal C}$}({\cal F}^\dag,\mbox{\boldmath $\cal F$}) \,,
\end{eqnarray}
where now $\cal F$ contains the full set of (sub-)CFFs and $\mbox{\boldmath $k$}^{-1}$ and $\mbox{\boldmath $K$}$ are matrices of dimensions $n$. For cross section measurements with an unpolarized nucleon target and a polarized electron beam \cite{Defurne:2015kxq,Jo:2015ema}
the information which can be extracted  are the first (weighted) harmonic of the beam helicity dependent
cross section $\Delta(\phi) = \frac{1}{2}(d\sigma^\uparrow - d\sigma^\downarrow)/d\xB\Q^2dtd\phi $,
\begin{subequations}
\label{BSDwBSSw-appr-tw2}
\begin{equation}
\BSD{} \approx \frac{{\cal N}}{\xB c^{\cal P}_0  t}
\frac{\xB  S_{++}(1)}{y} \Im{\rm m}  {\cal C}_{\rm unp}^{{\cal I}}({\cal F})
\label{BSDw0-appr-tw2}
\end{equation}
and  zeroth and first (weighted) harmonics of the unpolarized cross section $\Sigma(\phi) = \frac{1}{2}(d\sigma^\uparrow + d\sigma^\downarrow)/d\xB\Q^2dtd\phi $,
\begin{equation}
 \BSS{0} \approx \frac{{\cal N}}{\xB c^{\cal P}_0  t} \left[
 \frac{c^{\rm BH}_{0,{\rm unp}}}{\left(1+\epsilon ^2\right)^2}+ \frac{\xB  C_{++}(0)}{y} \Re{\rm e}  {\cal C}_{\rm unp}^{{\cal I}}({\cal F})+
 \frac{2\xB^2 c^{\cal P}_0  t}{\Q^2}  \frac{2-2 y + y^2+\frac{\epsilon ^2}{2} y^2}{1+\epsilon^2}\,
{\cal C}_{\rm unp}^{\rm VCS} ({\cal F},{\cal F}^\ast)
 \right],
\end{equation}
\begin{equation}
 \BSS{} \approx \frac{-{\cal N}}{\xB c^{\cal P}_0  t} \left[
 \frac{c^{\rm BH}_{1,{\rm unp}}}{\left(1+\epsilon ^2\right)^2}+ \frac{\xB  C_{++}(1)}{y}\Re{\rm e}  {\cal C}_{\rm unp}^{{\cal I}}({\cal F})+
 \frac{2w_1\xB^2 c^{\cal P}_0  t}{\Q^2} \frac{2-2 y + y^2+\frac{\epsilon ^2}{2} y^2}{1+\epsilon^2}\,
{\cal C}_{\rm unp}^{\rm VCS} ({\cal F},{\cal F}^\ast)
 \right],
\end{equation}
\end{subequations}
allowing us to map directly the experimental data to the imaginary and real part of the CFF combination ${\cal C}_{\rm unp}^{{\cal I}}({\cal F})$ and the bilinear form ${\cal C}_{\rm unp}^{\rm VCS} ({\cal F},{\cal F}^\ast)$. For precise definition of various symbols see Ref.\ \cite{Belitsky:2012ch}.

In the practical application of the strategy (\ref{F2A-gen}) one might straightforwardly calculate the sub-CFFs $\mbox{\boldmath $\cal F$}(\mbox{\boldmath $X$};F_1,F_2|\mbox{\boldmath ${\cal G}$})$, \emph{i.e.}, their mean and variance, as function of the experimental measurements, the form factor parametrization,
and some given values of the remaining CFFs $\mbox{\boldmath ${\cal G}$}$, whose fixed values can be varied in some model-dependent  or theory-inspired  region. Thus, the uncertainty of the extracted sub-CFFs $\mbox{\boldmath $\cal F$}(\mbox{\boldmath $X$};F_1,F_2|\mbox{\boldmath ${\cal G}$})$ contains essentially two main sources: the experimental uncertainty of the DVCS  and the form factor measurements, including also the uncertainties of QED corrections,
and the estimated uncertainty of the unknown CFF contributions. The strategy (\ref{X2C-gen}) has the advantages that within this assumption the uncertainties of the extracted linear and bilinear CFF combinations are governed only by the experimental uncertainties. However, some care is needed to have control over the uncertainties that are induced by the assumed approximation. A consistency check between the two methods is provided by evaluating the set of (approximate) equalities
\begin{equation}
\mbox{\boldmath ${\cal C}$}({\cal F}|F_1,F_2) \approx \mbox{\boldmath ${\cal C}$}(\mbox{\boldmath $\cal F$}(\mbox{\boldmath $X$};F_1,F_2|\mbox{\boldmath ${\cal G}$})|F_1,F_2)\,,
\quad
\mbox{\boldmath ${\cal C}$}({\cal F}^\dag,\mbox{\boldmath $\cal F$})  \approx
\mbox{\boldmath ${\cal C}$}(\mbox{\boldmath $\cal F$}^\dag(\mbox{\boldmath $X$};F_1,F_2|\mbox{\boldmath ${\cal G}$}^\dag),\mbox{\boldmath $\cal F$}(\mbox{\boldmath $X$};F_1,F_2|\mbox{\boldmath ${\cal G}$}))\,,
\end{equation}
which have to be (almost) independent on the values of $\mbox{\boldmath ${\cal G}$}$.

\subsection{Quasi model-independent method based on reference GPD models}
\label{sec:model-independent}

Another strategy, based on the method of least squares fitting, was suggested in Ref.~\cite{Guidal:2008ie} and
utilized to locally extract GPD information in LO and LT approximation (\ref{eq:CFF}) from experimental measurements
\cite{Guidal:2009aa,Guidal:2010ig,Guidal:2010de}.  Here, the parameters are eight multipliers, where their values are normalized to reference GPD predictions of the VGG code, reading here in our notation as
\begin{equation}
\label{VGGparameter}
\Im{\rm m}{\cal F}(\xB,t,\Q^2) = a^{\Im{\rm m}}_{\cal F} \Im{\rm m}{\cal F}^{\rm VGG}(\xB,t,\Q^2)\,,
\quad
\Re{\rm e}{\cal F}(\xB,t,\Q^2) = a^{\Re{\rm e}}_{\cal F} \Re{\rm e}{\cal F}^{\rm VGG}(\xB,t,\Q^2)\,,
\end{equation}
where ${\cal F}^{\rm VGG}$ are predicted by the VGG code via the convolution formula (\ref{eq:CFF}) and the multipliers $a^{\Im{\rm m}}_{\cal H},\cdots ,a^{\Re{\rm e}}_{\widetilde{{\cal E}}}$ are the eight new model-independent fitting parameters.  Further reduction might be made by setting some CFFs to zero, \emph{e.g.}, the imaginary part of the helicity non-conserved CFF $\widetilde{\cal E}$ whose smallness is suggested by the GPD findings in the DVMP analyses of $\pi^+$ electroproduction  \cite{Bechler:2009me}.
Often the method was/is employed to an incomplete set of observables, \emph{i.e.}, the fitting problem is ill-posed and to reach convergency one constrains the seven parameters $|a^{\Im{\rm m}}_{\cal F}| \lesssim 5$ or so, including the more unreliable model estimates $|a^{\Re{\rm e}}_{\cal F}| \lesssim 5$ for the real part of CFFs. This method has been recently extended by  generating an ensemble of fits with randomly distributed start values, where it is hoped that in this statistical manner also ill-posed fitting problems can be managed \cite{Boer:2014kya}.

\subsection{Comparison of results}

Various strategies to locally extract CFFs from the almost over-complete DVCS asymmetry measurements of the HERMES collaboration  were studied in
Ref.\ \cite{Kumericki:2013br}. Namely, one-to-one mapping, stepwise regression, and least squares fitting methods.  The quasi model-independent method, see Sec.\ \ref{sec:model-independent},
utilizes all available harmonics, \emph{i.e.}, together with the stronger hypothesis that all observables are describable in the hand-bag approach noise is used to stabilize the fitting procedure.  The results for $\{\Im{\rm m}{\cal H},\Im{\rm m}\widetilde{\cal H},\Re{\rm e}{\cal H}$\} of Ref.\ \cite{Boer:2014kya} are in good agreement with our findings \cite{Kumericki:2013br}. Note that in Ref.\ \cite{Kumericki:2013br} all local CFF values were reported, including the negative one for $\widetilde{\cal H}$ at $\xB=0.08$, $\Q^2=1.9\, \GeV^2$, and $-t=0.03\, \GeV^2$, however, we concluded that only the sub-CFFs $\Im{\rm m}{\cal H}$, $\Re{\rm e}{\cal H}$, $\Im{\rm m}\widetilde{\cal H}$, and maybe $\Im{\rm m}{\cal E}$ are accessible from HERMES measurements.

\begin{figure*}[t]
\centerline{\includegraphics[scale=0.44]{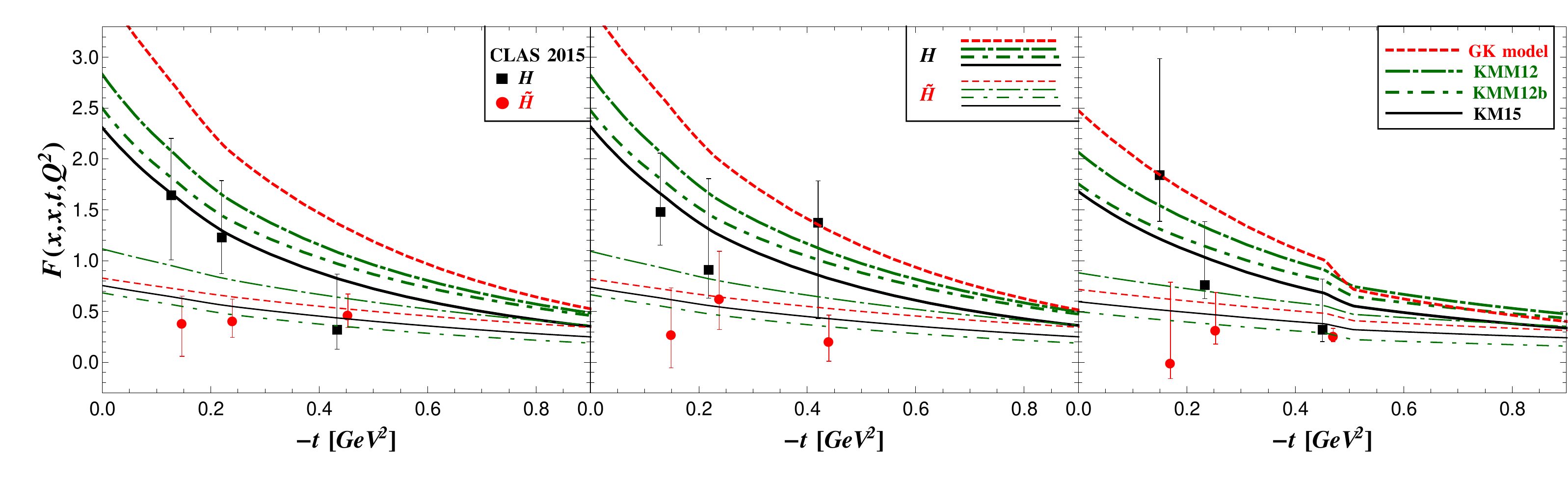}}
\vspace*{8pt}
\caption{GPD $H$ (filled squares, slightly shifted to the l.h.s., and thick curves) and $\widetilde{H}$ (filled circles, slightly shifted to the r.h.s., and thin curves) from DVCS asymmetry measurements in three $\{\xB, \Q^2\}$-bins versus $-t$ \cite{Pisano:2015iqa}. They are locally extracted by the quasi model-independent method (squares and circles) \cite{Pisano:2015iqa}, predicted by the \GK\ model (dashed),   \KM{M12} (dash-dotted)  fit, \KM{M12b} (shorter dash-dotted) fit, and from the new \KM{15}\ fit (solid).
}
\label{fig:CLAS15-asym2GPDs}
\end{figure*}
The new CLAS beam spin asymmetry $A_{\rm LU}$ and longitudinally polarized target spin asymmetry $A_{\rm UL}$  and $A_{\rm LL}$ measurements \cite{Pisano:2015iqa}
were also analyzed with the quasi model-independent tool,
shortly described in Sec.\ \ref{sec:model-independent}, were results for the imaginary parts of CFFs ${\cal H}$ and $\widetilde {\cal H}$ were presented
in Fig.\ 25 of Ref.\ \cite{Pisano:2015iqa}. In contrast to the HERMES analysis, there are only three asymmetries that are sensitive to two linear combinations
(\ref{C_unp}) and (\ref{C_LP}) of the imaginary part of CFFs and one to the real part of Eq.\ (\ref{C_LP}), where in addition the normalization of asymmetries, \emph{i.e.}, the unpolarized cross section that is dominated by the Bethe-Heilter amplitude, is nonlinearly governed by all CFFs. We emphasise that CLAS data in the DVCS region do not indicate (apart from noise) the appearance of higher harmonics and that there might be some tension among longitudinal target spin asymmetry measurements from HERMES and CLAS collaborations.  The results from the quasi model-independent approach have naturally large error bars, which  provide a statistical representation of both the propagated experimental uncertainties and the model-dependent constraints on the non-extractable degree of freedom.  As we would have expected, the locally extracted results in the DVCS kinematics are predicted from our older \KM{M12b}\ and  new \KM{15} fit (solid curves), shown for three DVCS bins in Fig.~\ref{fig:CLAS15-asym2GPDs} while the \GK\ (dashed) model prediction for GPD $H$ is too large. Note that all models have a rather similar GPD $\widetilde{H}$ prediction. The $t$-dependence of GPD $\widetilde{H}$ comes out flatter than for GPD $H$, which has a natural explanation in GPD models that incorporate form factor constraint. Namely, from the common GPD modeling and the fact that the axial nucleon form factor possesses a flatter $t$-dependence than the Pauli form factor, the above interpretation can be considered a model-dependent GPD prediction and only by means of the GPD framework one might loosely link the CFF $\widetilde{\cal H}$ to the axial-charge distribution, quantified by the
axial form factor.

\begin{figure*}[t]
\centerline{\includegraphics[scale=0.38]{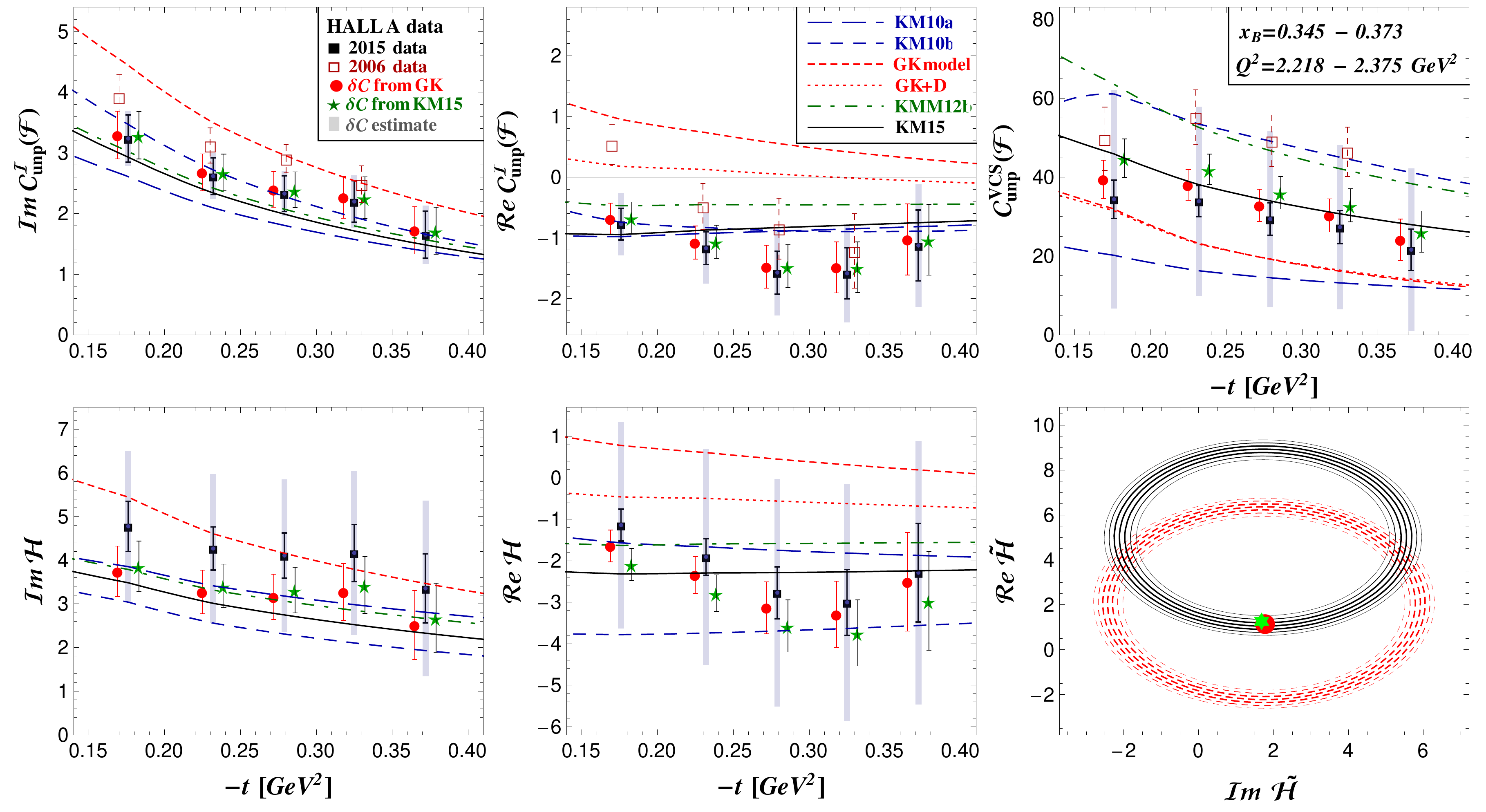}}
\vspace*{8pt}
\caption{Upper row: imaginary part [left panel] and real part [middle panel] of the linear CFF combination ${\cal C}_{\rm unp}^{\cal I}({\cal F})$, defined in Eq.\ (\ref{C_unp}), as well as of the bilinear CFF combination ${\cal C}_{\rm unp}^{\rm VCS}({\cal F}^\ast_{\rm unp},{\cal F})$ [right panel] are mapped from the Hall A cross section measurements  \cite{Defurne:2015kxq} for Kin3 kinematics (\ref{HallA-KINs}) within the Hall A approximation (filled squares) and utilizing the CFF ${\cal E}$ and $\widetilde {\cal E}$ values from the \GK\ model (filled circles)  and the \KM{15}\ fit (stars). The bands show the model uncertainties of power suppressed contributions in the range $|\Im{\rm m}{\cal E},\Im{\rm m}\widetilde{\cal H}|\le 2$ and $|\Re{\rm e}{\cal E},\Re{\rm e}\widetilde{\cal H}|\le 4$.  The empty squares sketch the first Hall A measurements \cite{Munoz_Camacho:2006hx} at slightly different kinematics.
Lower row: possible region for $\Im{\rm m}{\cal H}$ [left panel] and $\Re{\rm e} {\cal H}$ [left panel] within the same CFF constraints as well as the 1-$\sigma$ ellipses for the $\widetilde{\cal H}$ constraint  at $t=-0.232\, \GeV^2$ [right panel] for fixed CFF ${\cal E}$ and $\widetilde {\cal E}$ values  taken from the  \GK\ model [dashed curves] and the \KM{15}\ fit [solid curves].
Model predictions  are shown as curves, see Fig.\ \ref{fig:F2}.
The dotted curves show a RDDA based GPD $H$, taken from the \GK\ model, and the $D$-term form factor parametrization of Ref.\ \cite{Pasquini:2014vua}.
\label{fig:HallA15-CFFs}
}
\end{figure*}

As explained in Sec.\ \ref{sec:mapping},
the information which can be extracted from the new Hall A  \cite{Defurne:2015kxq} and CLAS \cite{Jo:2015ema} cross section measurements  are the imaginary and real parts of the CFF combination ${\cal C}_{\rm unp}^{{\cal I}}({\cal F})$ and the bilinear form ${\cal C}_{\rm unp}^{\rm VCS} ({\cal F},{\cal F}^\ast)$, see Eq.\ (\ref{BSDwBSSw-appr-tw2}), while extraction of sub-CFFs is an ill-posed problem.  We utilized the method (\ref{BSDwBSSw-appr-tw2}), where uncertainties were added in quadrature and propagated via the transformation of the covariance matrix, see \cite{Kumericki:2013br}.
Note that in the propagation of experimental uncertainties we ignored the fact that systematic ones are asymmetric; however, for the purpose of an illustration at twist-two level this is not an important issue.
Our results, shown as filled squares in the upper row of Fig.\ \ref{fig:HallA15-CFFs} coincide for the imaginary and real part of the linear combination ${\cal C}_{\rm unp}^{\cal I}$ and for the bilinear form ${\cal C}^{\rm VCS}$  with those that were obtained by the Hall A collaboration by fits to the $\phi$--dependent cross sections, see Fig.\ 22 of Ref.\ \cite{Defurne:2015kxq}, where the net uncertainties are  dominated by the systematic errors. The same method applied to the previous Hall A data \cite{Munoz_Camacho:2006hx}, shown by empty squares, reveals some differences between the data sets (here we assume that these CFF combinations are smooth functions that vary only slightly within the small kinematical difference).   However, as we will explain now, the adopted approximation implies an additional uncertainty and so the net errors of these results might be underestimated.

On the other hand we might map the experimental data to the CFF ${\cal H}({\cal E },\widetilde{\cal H})$ and the bilinear form ${\cal C}^{\rm VCS}({\cal F}^\ast,{\cal F})$ by utilizing the ``exact'' formulae for the restricted set of four CFFs. As expected, the extracted mean values for the linear combination
$$
\Im{\rm m}{\cal C}_{\rm unp}^{\cal I}(\Delta|{\cal E}=\widetilde {\cal H}=0) \mbox{ \ and \ }
\Re{\rm e}{\cal C}_{\rm unp}^{\cal I}(\Sigma|{\cal E}=\widetilde {\cal H}=0)
$$
only slightly deviate from the findings within the direct extraction procedure in which power suppressed corrections are neglected. In average we find that the results deviate on the 0.2--0.3\% level where the maximal deviation does not exceed 0.5\% and 8\% for the imaginary and real parts, respectively. Taking values for the unconstrained CFFs from the \GK\ model prediction (filled circles) or our \KM{15} fit (stars) will only slightly change the mean values for this linear CFF combinations.
In particular for the extraction of the bilinear form power suppressed corrections play a more important role and, thus, the two different methods provide results where in average the differences are  $\sim 7 \%$, however, can be up to $\sim 15 \%$ in some bins. The results where the unconstrained
CFFs ${\cal E}$, $\widetilde{\cal H}$, and $\widetilde{\cal E}$ are set to the \GK\ model and the  \KM{15} fit findings are shown as filled circles and stars, respectively.
For both choices the values for $\widetilde{\cal H}$ are very similar and since the sensitivity to ${\cal E}$ is much more suppressed, we find rather similar results which are also very similar to those obtained in the ${\cal E}=\widetilde {\cal H}=0$ case.
However, varying the unconstrained CFFs in the region
\begin{equation}
\label{EtH-variation}
|\Im{\rm m}{\cal E},\Im{\rm m}\widetilde{\cal H}|\le 2 \mbox{ \ and \ } |\Re{\rm e}{\cal E},\Re{\rm e}\widetilde{\cal H}|\le 4
\end{equation}
generates for the real part of the linear combination and  for the bilinear form a larger variation of the extracted mean values inside the gray rectangles, which reflects the importance of power suppressed contributions. Again the differences are much more pronounced for the extracted values of the bilinear CFF coefficient ${\cal C}_{\rm unp}^{\rm VCS}$. Taking into account this additional uncertainty, we can conclude that our previous \KM{10b} (\cite{Kumericki:2011zc}, dashed curves) and \KM{M12b} (dash-dotted curves) fits in which only the relative $\phi$-modulation of the superseded dataset \cite{Munoz_Camacho:2006hx} was utilized predicts the locally extracted values of the new data \cite{Defurne:2015kxq} while the new global \KM{15} fit (solid curves) is entirely consistent with the locally extracted values. Compared to the older data set (empty squares), the \GK\ model prediction (short dashed curves and dotted curves that adds a D-term contribution from Ref.\ \cite{Pasquini:2014vua} ) gets more disfavored with respect to the new data set (filled squares).

Clearly, if one asks for the values of the CFF ${\cal H}$ the uncertainties depend much on the constraints for the remaining CFFs. This is illuminated in the first two panels of the lower row in Fig.\  \ref{fig:HallA15-CFFs} where due to the variation (\ref{EtH-variation}) the mean values of ${\cal H}({\cal E}=0,\widetilde{\cal H}=0)$  (filled squares) can move in the region of the rectangles. The findings for $\Im{\rm m}{\cal H}$ would be consistent with the values from the RDDA based \GK\ model, however, taking values for $\widetilde{\cal H}$ and ${\cal E}$ from the \GK\ model (circles) or the \KM{15}\ fit (stars) we realize that $\cal H$ is consistent with the \KM{15}\ fit and to some extent with our model predictions.   In the right panel of the lower row we show the correlation between the real and imaginary part of the CFF $\widetilde{\cal H}$ within the 1-$\sigma$ band and with remaining CFFs fixed by the \GK\ model (dashed curves) or the \KM{15}\ fit (solid curves). The model values for $\widetilde{\cal H}$ are again displayed as filled circle and star and one realizes that the  \GK\  model violates this constraint while the \KM{15} fit respects it.

\begin{figure*}[t]
\centerline{\includegraphics[scale=0.38]{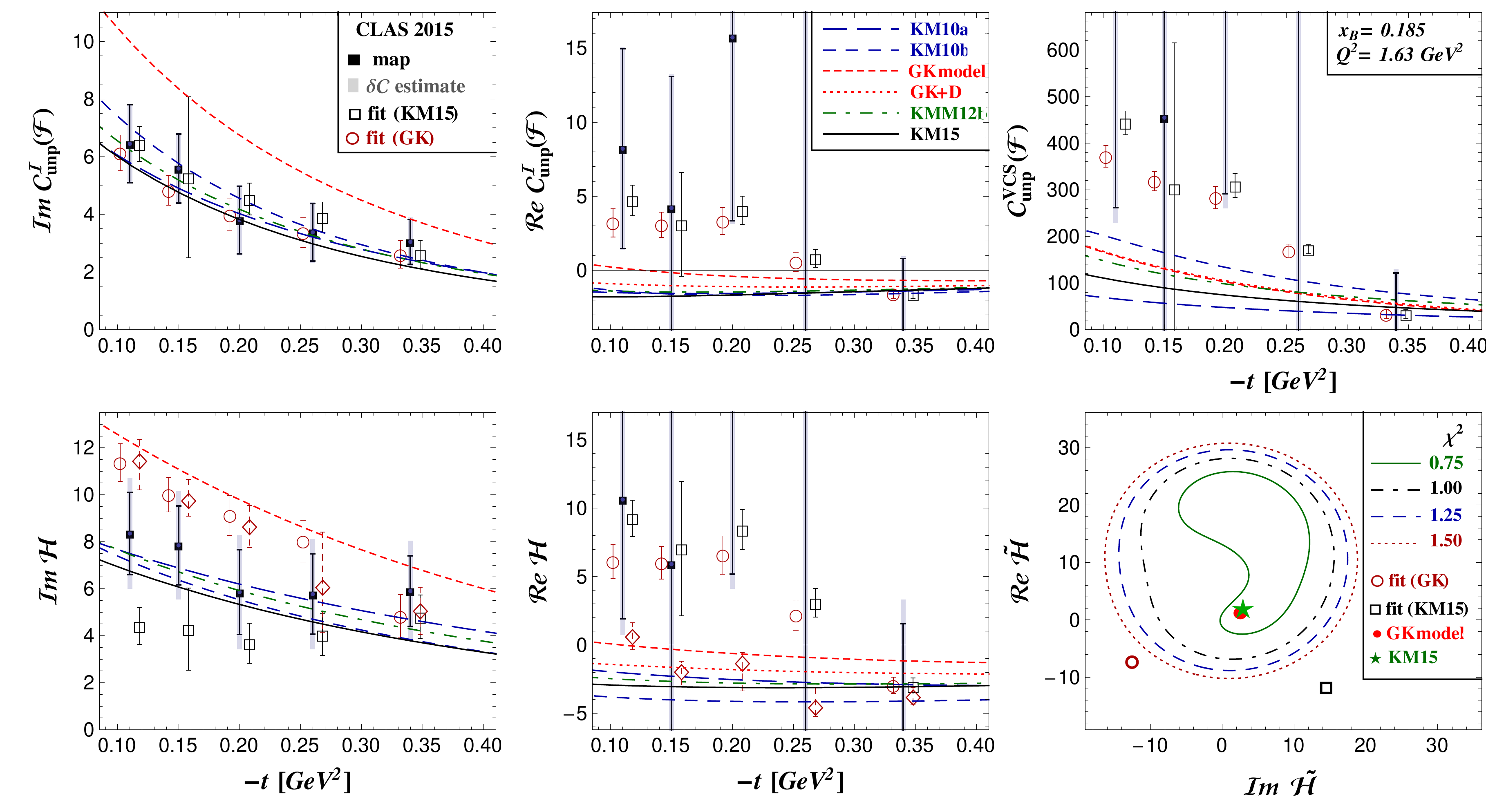}}
\vspace*{8pt}
\caption{Imaginary  [upper left panel] and real [middle panel]  parts of linear CFF combination ${\cal C}^{\cal I}_{\rm unp}$ as well as the bilinear combination ${\cal C}^{\rm VCS}_{\rm unp}$ [upper right panel]
for the $\{\xB=0.185, \Q^2=1.63\; \GeV^2\}$-bin are locally extracted  from CLAS data \cite{Jo:2015ema} by a map (filled squares), where rectangles show the variation of mean values by utilizing the constraint (\ref{EtH-variation}),  and from a constrained four parameter, $\Im{\rm m} {\cal H}$, $\Im{\rm m} \widetilde{\cal H}$, $\Re{\rm e} {\cal H}$, $\Re{\rm e} \widetilde{\cal H}$, fit with start values from the \GK\ model (empty circles) and \KM{15}\ fit (empty squares).
Model predictions are shown as curves and are described in Fig.\ \ref{fig:HallA15-CFFs}.
\label{fig:CLAS15vsCFFs}
}
\end{figure*}

The CLAS collaboration used the quasi model-independent tool and tried to extract the CFF ${\cal H}$ from the cross section measurements \cite{Jo:2015ema}, where only the two CFFs $\cal H$ and $\widetilde{\cal H}$ were taken as four fitting parameters. Apart from noise effects this is essentially a three parameter fit problem, see Eq.\ (\ref{BSDwBSSw-appr-tw2}), where, however, four constrained parameters were used. We utilized a similar method, shown for the $\{\xB=0.185, \Q^2=1.63\; \GeV^2\}$-bin in Fig.\ \ref{fig:CLAS15vsCFFs} by the empty circles and squares. In the lower panel row we show that the resulting values, in particular for $\Im{\rm m} {\cal H}$ (left panel), might depend on the initial values and we might reproduce the CLAS result (see empty rhombi) if we take for the initial values the \GK\ model predictions, however, we might find another solution if we utilize the \KM{15}\ result (empty squares) for the start values. Compared to our mapping procedure, providing us ${\cal H}({\cal E},\widetilde{\cal H})$ (filled squares and rectangles), the uncertainties from the fitting procedures are underestimated. The reason is that the unconstrained degrees of freedom tend to be on the boundary, see $t=-0.11\ \GeV^2$ example in the lower right panel,  and thus the error propagation is not reliable. Such large values for $\widetilde{\cal H}$ can also be considered as inconsistent with the findings from the new CLAS asymmetry measurements, described above. Plugging our CFF fit results into the CFF combinations, see upper panel row, provides results that are consistent with our mapping procedure, however, the uncertainties from the fits might still be underestimated. Our model predictions and fit results are consistent with the CFF values from the map. In contrast to the CLAS statement, we conclude that the locally extracted ${\cal H}$ values disfavour a RDDA based GPD model (here the \GK\ model).

\section{Conclusions}

We compared model predictions with the new DVCS measurements from the CLAS and Hall A collaborations where we utilize a (weighted) Fourier transform as filter to strengthen the discriminating power of predictions.  We presented a new global DVCS fit \KM{15} showing that superseding 2006 Hall A cross section data with new one relieves some tensions, but description of first cosine harmonic of cross section remains imperfect. We showed that our global fit predictions \KM{10b}\ and \KM{M12b}, which ignored the normalization of the first Hall A cross section measurements,  predict the new measurements reasonably well while a RDDA based model, \emph{e.g.}, the \GK\ model, becomes more disfavored by the new data sets. These conclusions are partially in disagreement with the first analyses of the data by collaborations themselves. In particular, the new cross section measurements support the previous findings that even in fixed target kinematics such models provide a GPD $H$ that is too large. We further provided within our simple GPD ansatz from previous fits a good description of the new global DVCS data set and showed that the locally extracted CFF values are consistent with the global description. This analysis includes the GPD $\widetilde{H}$ which is rather compatible with RDDA based model predictions with respect to normalization and $t$-dependence which comes out flatter than for GPD $H$. Due to the new cross section measurements the indication that the GPD $H$ has a rather small skewness effect becomes more reliable.

\begin{acknowledgement}
We like to thank M.~Guidal for a comparison of local fitting results that were obtained from our and his fitting analysis and we are indebted to H.~Avakian,  M.~Garcon, and S.~Niccolai for discussions.
This work has been supported in part by the Croatian
Science Foundation under the project number 8799, Croatian Ministry of Science, Education and Sport (MSES) under the NEWFELPRO grant agreement
no.\ 54, and by the NRF of South Africa under CPRR grant no.~90509.
\end{acknowledgement}

% For bibliography use \cite{RefJ}
% BibTeX or Biber users please use (the style is already called in the class, ensure that the "woc.bst" style is in your local directory)
% \bibliography{name or your bibliography database}
%\bibliographystyle{woc}
%\bibliography{../../../../myTeXinput/veroefli,../../../../myTeXinput/referenc}

%\input{KKumericki-DMueller.bbl}

%\begin{thebibliography}{}
%Journal Author, Journal \textbf{Volume}, page numbers (year)
%Book Author, \textit{Book title} (Publisher, place, year) page numbers
% etc
%\end{thebibliography}

\end{document}